\documentclass{article}
\usepackage[utf8]{inputenc}
\usepackage{amsmath}
\usepackage{amsfonts}
\usepackage{amssymb}
\usepackage{amsthm}
\usepackage{stackrel}
\usepackage{graphicx}
\usepackage{tikz}
\usepackage{xcolor}
\usepackage[export]{adjustbox}
\usetikzlibrary{arrows,positioning, shapes} 
\usepackage{float}
\usepackage[utf8]{inputenc}
 \usepackage[T1]{fontenc}
\usepackage{url}
\usepackage{subfig}
\usepackage{float}

\title{Depth for samples of sets with applications to testing equality in distribution of two samples of random sets}
\author{\footnote{$^1$ University of Split, Croatia  ($^{\ddag}$ email: vgotovac@pmfst.hr)} Vesna Gotovac \DJ oga\v{s}$^1$}

\begin{document}
\maketitle
\begin{abstract}
    This paper introduces several depths for random sets with  possibly non-convex realisations, proposes ways to estimate the depths based on the samples and compares them with existing ones. The depths are further applied for the comparison between two samples of random sets using a visual method of DD-plots and statistical testing. The advantage of this approach is identifying sets within the sample that are responsible for rejecting the null hypothesis of equality in distribution and providing clues on differences between distributions. The method is justified using a simulation study and applied to real data consisting of histological images of mastopathy and mammary cancer tissue.
\end{abstract}
\section{Introduction}
Random sets play an important role as a mathematical tool for understanding the geometry of many phenomena in science.
Some of the many examples are describing the shape of different types of tissue in medicine  \cite{hermann:2015}, the spatial arrangement of the plants in the ecosystem \cite{moeller:2010}, the microstructure of the materials \cite{neumann:2016}.
The rich theory of random sets is  discussed in 
\cite{matheron:1975}, \cite{molchanov:2005}, \cite{serra:1982}.

In the applications mentioned above, the random set processes are obtained as a union of the components of different shapes.
Although it is advantageous to have a concrete random set model of the components, it is not indispensable when we only need to compare the processes based on some specific features or we just need to detect the outliers with respect to some property within the samples.
Furthermore, in many applications, the shape of those components may be too complex to describe with a simple enough model.

In this paper, we focus on the non-parametric analysis of the distribution of the random sets via statistical depths.

Statistical depth is a powerful tool for this purpose. Its introduction aimed to overcome the lack of order in $\mathbb R^d$ and provides centre outward ordering of the elements in the support of the distribution. In this way, we can generalise the concepts of quantiles and ranks.
It has been used for various purposes and in numerous fields for the investigation of outlieingness as well as location, scale, bias, skewness and kurtosis of the sample's distributions. For the comparison of the distributions, a graphical tool used is a DD-plot \cite{ddplot1}.
The halfspace depth was first to be introduced in \cite{Tukey} where the depth of a multivariate point is defined as the minimum
probability mass carried by any closed halfspace having that point on the boundary. 

The other well-known concept of multivariate depth is a simplicial depth where the depth of a point was defined as a probability of the point belonging to a simplex whose vertices are i.i.d. sample from a given distribution  (see \cite{liu:1990} and \cite{cascos:2010}).
Many other multivariate depth concepts are summarised and classified based on their properties and constructive approaches in \cite{zuo:2000}.

Generalisation of the multivariate depth concept to functional data was challenging since the sample of functions is meagre set in the functional space causing most of the sample functions to have depth zero. The first depth for functional data was integrated depth (see \cite{fraiman:2001}), where the univariate depths of the function values were averaged among the domain, followed by the band depth (\cite{lopez:2009}) defined as the probability of function being sandwiched between pointwise minima and maxima of fixed number of i.i.d. random curves and infimal depth (\cite{mosler:2013},\cite{gijbels:2015}) calculated as the infimum over univariate depths of the functional values for the fixed arguments. Many other concepts of functional depth where the functions are 
classified using their shape can be found in \cite{gijbels:2017}.

Sets can be represented as functions (see \cite{molchanov:2005}). The most obvious representation is obtained using an indicator function. Some other representations more frequently used in image analysis include distance and signed distance functions. If the sets are convex bodies, then the most natural representation is by their support functions, however, this is not the setting in this paper.

Different concepts of depth for random convex sets and their samples are well studied in \cite{outliers_anz}. However, in many applications, we encounter sets that are non-convex and have a complex structure which is hard to describe using a particular model.

Therefore, in this paper, we focus on discussing and summarising possible ways to define the depth for general random sets. The aim is not to cover all the theoretical possibilities for defining depth for random-non convex sets, but rather concentrate on the ones that can be easily estimated from the samples of random sets in real time. Also, we want to compare the depths to get an insight into how they order the set realisations, e.g. what specific features are taken into account when the ordering is done.

For the general random sets concept, \cite{whitaker:2013} have already extended the functional band depth approach by representing sets by their indicator functions.  This approach leads to the construction of the depth based on the intersections and unions of the i.i.d. sets.

We also propose a new test for equality in the distribution of two samples of random sets using statistical depths.

Test for equality in the distribution of random non-convex sets has already been proposed in \cite{gotovac:2019}. There the permutation test based on $\mathfrak N$-distances was used, where the negative definite kernel employed was the area of symmetric difference between sets. However, this procedure does not provide clues on the reasons for the rejection of the null hypothesis. Here we address this problem by using DD-plots for the graphical representation of the differences between the distributions of two samples. Also, the testing procedure developed in this paper, in the case of rejection of the null hypothesis, provides the sets in the sample responsible for the rejection.

In this research, we concentrate on the planar setting because of easier visualisation. However, all the presented methodologies can be easily modified for the more dimensional case.

The remainder of the paper is organised as follows. Section 2 presents the theory of random sets needed for understanding the paper and states the desirable properties of statistical depth for random sets.
Section 3 summarises different depth concepts for random sets and compares them.
In section 4  we discuss ways to compare distributions of random sets using DD-plots and present the test for equality in the distribution of samples of random sets. Section 5 is dedicated to the simulation study where we investigate the power of the proposed test in various cases of alternative hypothesis. In Section 6 we apply the proposed procedure to histological images of mammary cancer and mastopatic tissue data. The concluding remarks are summarised in the Discussion.

\section{Theoretical background}
Let $(\Omega, \mathfrak{F}, \mathbf{P})$ a  probability space and $\mathcal{F}$ the family of closed sets in $\mathbb R^d.$
 A random closed set is a map $\mathbb X: \Omega \mapsto \mathcal{F}$ such that for every compact set $K$ in $\mathbb{R}^d$,
$$
\{\omega: \mathbb X \cap K \neq \emptyset\} \in \mathfrak{F} .
$$

A set $F \in  \mathcal F$ belongs to the support of random set $\mathbb X$ if $\mathbb X$ with a positive probability belongs
to any open neighbourhood of $F$ in the Fell topology.

A sequence $\left\{F_n, n \geqslant 1\right\}$ converges to $F \in \mathcal{F}$ in the Fell topology if $F_n \cap G \neq \varnothing$ for all sufficiently large $n$ and any open set $G$ that hits $F$, and $F_n \cap K=\varnothing$ for all sufficiently large $n$ and any compact set $K$ that misses $F.$ (see \cite{molchanov:2005}). 

For arbitrary closed sets $F_1, F_2 \in \mathcal F$ we define its Minkowski sum by
$$
F_1+F_2=\operatorname{cl}\left\{x+y: x \in F_1, y \in F_2\right\}, 
$$
where $\operatorname{cl}(F)$ denotes the closure of set $F.$
 Specially, $F+a=\{x+a: x \in F\}$, for $a \in \mathbb{R}^d$.
 For an arbitrary $a\in \mathbb R$ we define $aF=\{ax: x\in F\}.$

A random vector $\xi$ in $\mathbb{R}^d$ is a selection of $\mathbb{X}$ if $P(\xi \in \mathbb{X})=1$. The family of all integrable selections of $\mathbb{X}$ is denoted by $L^1(\mathbb{X})$.

A random closed set $\mathbb{X}$ is integrable if the family $L^1(\mathbb{X})$ is non-empty. For integrable random closed sets $\mathbb {X}$ selection (or Aumann) expectation is defined by
$$
\mathbb{E}(\mathbb{X})=\operatorname{cl}\left\{\mathbb{E}(\xi): \xi \in L^1(\mathbb{X})\right\}.
$$
The estimator of $\mathbb{E}(\mathbb{X})$ based on a sample $\mathbb X_1,\ldots , \mathbb X_n$ is given by $\widehat{\mathbb{E}(\mathbb{X})}=\frac{1}{n}\left( \mathbb X_1+\ldots+ \mathbb X_n\right)$ (see \cite{molchanov:1998}).

Let  $\mathbb {X}$ be a random closed set. Its depth function $D(\cdot, \mathbb{X})$ is a function of a closed set $F$ taking values in $\left[0,1\right]$. If $F$ does not belong to the support of $\mathbb{X}$ $\mathrm{D}(F, \boldsymbol{X})=0.$ ,

 Following generalisation of the properties of depths for random vectors were proposed in \cite{outliers_anz} for the case of the depths for random sets:
\begin{itemize}
    \item[(D1)] Affine invariance:
$$
\mathrm{D}(A F+b, A \mathbb X+b)=\mathrm{D}(F, \mathbb X)
$$
for all non-singular $d \times d$ matrices $A$ and $b \in \mathbb{R}^d$.
\item[(D2)] Upper semicontinuity:$$
\mathrm{D}(F, \mathbb X) \geqslant \limsup _{n \rightarrow \infty} \mathrm{D}\left(F_n, \mathbb X\right),
$$
if $F_n \rightarrow F$ as $n \rightarrow \infty$ in the Fell topology on  $\mathcal{F}$.
\item[(D3)] If $\mathbb X=L$ a.s. for $L \in \mathcal{F}$, then $\mathrm{D}(F, L)=\mathbf{1}(F=L)$.
\end{itemize}

\section{Statistical depth for random sets}
In this section, we discuss several depths for general random sets and propose ways how to estimate them based on the sample.
\subsection{Depths based on representations of closed sets by functions}
The closed set $F$ can be represented as a function in several ways. The most obvious one is the representation via its characteristic or indicator function $1_F(x)=\left\{\begin{array}{ll} 1, &x \in F,\\ 0, & x \notin F.\end{array}\right.$ 

If we represent the set by the indicator function and use infimal functional depth (see \cite{mosler:2013}, \cite{gijbels:2015}), we obtain:
\begin{equation}
D_{inf}(F,\mathbb X)=\inf_{x \in \mathbb R^d} \{P(1_{\mathbb X}(x)\leq 1_{F}(x)), P(1_{\mathbb X}(x)\leq 1_{F}(x)\}.
\label{eq:D_inf}
\end{equation}
The estimation of depth $D_{inf}$ based on the sample is obtained by replacing the distribution of $\mathbb X$ with its empirical version.

When using functional band depth together with the indicator function representation of the set, we get a nice formulation of set depth in terms of random set notation
\begin{equation}
    D_{band}(F,\mathbb X)=P\left(\bigcap\limits_{i=1}^{n}\mathbb X_i \subseteq F \subseteq \bigcup\limits_{i=1}^{n}\mathbb X_i\right).
    \label{depth2}
\end{equation}
where $\mathbb X_1,\ldots,\mathbb X_n$ are i.i.d. random sets with the same distribution as $\mathbb X$ and $n \in \mathbb N.$ This depth was introduced in \cite{whitaker:2013} where its properties are also derived. 
It can be obtained that $D_{band}$ satisfies properties (D1)-(D3) (see \cite{outliers_anz}).

Suppose now we have a sample $\mathcal X=\{ X_1,\ldots,X_{M}\},$ i.e. the realisation of $M$ i.i.d. random sets $\mathbb X_1,\ldots, \mathbb X_M$ with the same distribution as  $\mathbb X.$ For chosen  $n\in \mathbb N, \ n<=M,$ the estimate of the depth (\ref{depth2}) is given by following U-statistics
\begin{equation}
D_{band,n}(F,\mathcal X)=\widehat{D_{band}(F,\mathbb X)}=\frac{1}{\binom{M}{n}}
 \sum_{1 \leqslant k_1<\cdots<k_n \leqslant M} \mathbf{1}\left(\bigcap_{i=1}^n X_{k_i} \subseteq F \subseteq \bigcup\limits_{i=1}^n X_{k_i}\right).
 \label{eq:D_band_est_s}
\end{equation}
For large $M$ and $n,$ $\binom{M}{n}$ can become extremly large to make the calculations of (\ref{eq:D_band_est_s}) too slow. In these cases, we can choose $s<\binom{M}{n},$ bootstrap from the sample $s$ samples of length $n$ $X^{(j)_1},\ldots,X^{(j)}_n, j=1,\ldots,s$ and obtain 
\begin{equation}
 D_{band,n,s}(F,\mathcal X)=\widehat{D_{band}(F,\mathbb X)}=\frac{1}{s}
 \sum_{j=1}^s \mathbf{1}\left(\bigcap_{1=1}^n X^{(j)}_i \subseteq F \subseteq \bigcup\limits_{i=1}^n X^{(j)}_{i}\right).
 \label{eq:D_band_est}
\end{equation}

The other representations of a closed set by the function that are common in image analysis are representations by the distance function or the signed distance function. Further on, we concentrate on the representation by the signed distance function.
The signed distance function of set $F\subset W$ is defined by $f_F: W \to \mathbb R$  as
\[f_F(x)=\left\{\begin{array}{ll}
d(x,F), & x \notin F,\\
-d(x,W\diagdown F) & x \in F,
\end{array}\right. \]
where $d(x,F)$ is a arbitrary distance from a point $x$ to a set $F.$
Now, we can identify each closed set $F$ with its signed distance function and use arbitrary functional depth for calculating the depths. These depths will be denoted by $D_{sign}.$ Note that this approach is similar to Example 6.1 in \cite{outliers_anz}.
Depths based on signed distance functions do not satisfy properties (D1)-(D3) in general since  those properties rely on the properties of the chosen distance $d.$
\subsection{Depth based on the distance }
Generalising the Type B depth from \cite{zuo:2000} for the case of random sets we come up with 
\[
D(F,\mathbb X)=\frac{1}{1+\mathbb E(d(F,\mathbb X))},\]

where $d: \mathcal F\times \mathcal{F} \to \mathbb{R}$ is a function that measures proximity from one set to another. Typical examples to be considered are the Hausdorff distance (further denoted by $D_{Haus}$) or the Lebesgue measure of the symmetric difference between sets (further denoted by $D_{Leb}$).
These depths can be estimated from the sample by replacing the distribution of the random set with its empirical version.

\subsection{Depth based on expectation of random set}
Suppose that we have functions $\mathcal E_{\alpha},\mathcal U_{\alpha}$ defined on some subset of a family of random sets taking values in $\mathcal F.$ 

A family $\left(\mathcal{U}_\alpha, \mathcal{E}_\alpha\right), \alpha \in(0,1]$, of pairs such that $\mathcal{U}_\alpha\subseteq \mathcal{E}_\alpha, \alpha \in \left\langle 0, \right]$is said to form a parametric family if, for each random closed set from its domain $\operatorname{set} \mathbb X, \mathcal{U}_\alpha(\mathbb X)$ is increasing and $\mathcal{E}_\alpha(\mathbb X)$ is decreasing (in the sense of set inclusions) as functions of $\alpha \in(0,1]$.
Let $F$ belong to the convex hull of the support of $\mathbb X$. Define a depth function of such $F$ as
$$
D(F, \mathbb X)=\sup \left\{\alpha \in(0,1]: \mathcal{U}_\alpha(\mathbb X) \subseteq F \subseteq \mathcal{E}_\alpha(\mathbb X)\right\}
.
$$
and $\sup \varnothing=0$.

Depth based on the functional of random sets satisfies properties (D1)-(D3) (see \cite{outliers_anz}).

Similar to Example 4.9. in \cite{outliers_anz} we consider 
\begin{equation} 
D_{exp}(F, \mathbb X)=\sup\{\frac{1}{m}: \mathbb E (\mathbb X_1\cap \ldots \cap \mathbb X_m)\subseteq F \subseteq \mathbb E (\mathbb X_1 \cup \ldots \cup \mathbb X_m) \}.
\label{eq:D_exp}
\end{equation}
The $\mathbb E (\mathbb X_1\cap \ldots \cap \mathbb X_m)$ and $\mathbb E (\mathbb X_1 \cup \ldots \cup \mathbb X_m)$ can be approximated from the sample of realisations is a following way.

suppose we have a sample $X_1,\ldots,X_n,$ $n \in \mathbb N.$ For chosen $S \in \mathbb N$ we resample $S$ samples with replacement form the original sample obtaining $X^(j)_i, i=1,\ldots,n, \ j=1,\ldots, S.$
Further, we estimate $\mathbb E (X_1\cap \ldots \cap \mathbb X_m)$ by $\widehat{\mathbb E ( X_1\cap \ldots \cap \mathbb X_m)}=\frac{1}{S}((X^{(1)}0_1\cap\ldots\cap X^{(1)}_m)+\ldots+(X^{(S)}_1\cap\ldots\cap X^{(S)}_m)$ and $\mathbb E (X_1\cup \ldots \cup \mathbb X_m)$ by $\widehat{\mathbb E ( X_1\cup \ldots \cup \mathbb X_m)}=\frac{1}{S}((X^{(1)}_1\cup\ldots\cup X^{(1)}_m)+\ldots+(X^{(S)}_1\cup\ldots\cup X^{(S)}_m).$ In this way we assure that $\widehat{\mathbb E ( X_1\cap \ldots \cap \mathbb X_m)}\subset\widehat{\mathbb E ( X_1\cup \ldots \cup \mathbb X_m)}.$ Then $\widehat{D(F, \mathbb X)}=\sup\{ m^-1: \widehat{\mathbb E (\mathbb X_1\cap \ldots \cap \mathbb X_m)}\subseteq F \subseteq \widehat{\mathbb E (\mathbb X_1 \cup \ldots \cup \mathbb X_m) \}}.$

One of the downsides of $D_{exp}$ is that it only can obtain values from discrete set $\{ 1,\frac{1}{2},\ldots, \frac{1}{m},\ldots\}.$
 It can be overcome by replacing $1/m$  in $(\ref{eq:D_exp})$ with $\lambda \in \left\langle 0,1\right\rangle $ and $X_m$ with $X_{N_{\lambda}}$ where $N_{\lambda}$ is a geometrical distributed random variable with parameter $\lambda.$ However, the estimation of this variant of depth is significantly more computationally expensive.
\subsection{Simplical depth}
We also consider a new approach for defining depths for random sets that could be natural in the models including Minkowski addition (see e.g. \cite{micheletti:2005}).

Let $\mathbb X$ be a random closed set, $\mathbb X_1,\ldots,\mathbb X_m$ a sequence of i.i.d. random sets with the distribution equal to $\mathbb X$ and $F$ an arbitrary closed set. 

Suppose that $F_1,\ldots, F_m$ are arbitrary closed sets. Denote by $conv(F_1,\ldots,F_m)=\{ p_1F_1+\ldots +p_mF_m: p_1,\ldots,p_m \geq 0, p_1+\ldots+p_m=1\},$ where $+$ stands for Minkowski addition of the sets.

We can define a simplicial depth \begin{equation}
D_{simp}(F,\mathbb X)=P(\{ \omega \in \Omega: \exists L, U \in conv(\mathbb X_1(\omega),\ldots,\mathbb X_m(\omega)) \text{ \ such that \ } \  L\subseteq F \subseteq U \}).
\label{eq:D_simp}
\end{equation}
Let us prove that the conditions $(D1)-(D3)$ hold.

$(D1)$
\begin{align*}
&D(AF+b,A\mathbb X+b)=P(\{ \omega \in \Omega: \exists U', L' \in conv(A\mathbb X_1(\omega)+b,\ldots,A\mathbb X_m(\omega)+b) \  L'\subseteq AF+b \subseteq U'\})=\\
&=P(\{ \omega \in \Omega: \exists L, U \in conv(\mathbb X_1(\omega),\ldots,\mathbb X_m(\omega)) \  AL+b\subseteq AF+b \subseteq AU+b\})=\\
&=P(\{ \omega \in \Omega:\exists L, U \in conv(\mathbb X_1(\omega),\ldots,\mathbb X_m(\omega)) \  L\subseteq F \subseteq U\})=\\
&=D(F,\mathbb X).
\end{align*}

$(D2)$

Suppose that $F_n \rightarrow F$ in Fell topology. Since for every finite measure $\mu$ and a sequence of measurable sets $(C_n)_n$  $\limsup\limits_{n}\mu(C_n)\leq \mu(\limsup\limits_{n}C_n),$ it holds that 
\begin{align*}
&\limsup\limits_{n} D(F_n,\mathbb X)=\\
&=\limsup\limits_{n}P(\{ \omega \in \Omega: \exists L_n, U_n \in conv(\mathbb X_1(\omega),\ldots,\mathbb X_m(\omega)) \  L_n\subseteq F_n \subseteq U_n\})\leq\\
&\leq P( \limsup\limits_{n}\{ \omega \in \Omega:\exists A_n, U_n \in conv(\mathbb X_1(\omega),\ldots,\mathbb X_m(\omega)) \  A_n\subseteq F_n \subseteq U_n\}\})=\\
&=P( \{ \omega \in \Omega: \exists L_{n_k}, U_{n_k} \in conv(\mathbb X_1(\omega),\ldots,\mathbb X_m(\omega)) \  L_{n_k}\subseteq F_{n_k} \subseteq U_{n_k}\})\leq\\
&\leq P( \{ \omega \in \Omega :\exists L_{n_k}, U_{n_k} \in conv(\mathbb X_1(\omega),\ldots,\mathbb X_m(\omega)) \  \bigcap\limits_k L_{n_k}\subseteq F_{n_k} \subseteq  \bigcup\limits_k U_{n_k}\})\leq \\
&\leq P( \{ \omega \in \Omega: \exists L_{n_k}, U_{n_k} \in conv(\mathbb X_1(\omega),\ldots,\mathbb X_m(\omega)) \  \bigcap\limits_k L_{n_k}\subseteq F \subseteq  \bigcup\limits_k U_{n_k}\})=\\
&=D(F,\mathbb X).
\end{align*}

$(D3)$

Suppose that $\mathbb X=B$ a.s., then $conv(\mathbb X_1,\ldots,\mathbb X_m)=B$ a.s. an the event $\{\exists L, U \in conv(\mathbb X_1,\ldots,\mathbb X_m) \  L\subseteq F \subseteq U\}$ equals to the event $\{ F=B \}.$ Therefore, $D(F,\mathbb X)=1(F=B).$

Suppose now we have a sample $\mathcal X=\{ X_1,\ldots,X_{M}\},$ i.e. the realisation of $M$ i.i.d. random sets $\mathbb X_1,\ldots, \mathbb X_M$ with the same distribution as  $\mathbb X.$ If we want to estimate the depth $D(X_i,\mathbb X),$ $i=1,\ldots,M,$ we can proceed using a bootstrap approach in a following way. First, we choose $m\leq M$ and $s \in \mathbb N.$ Then $s$ times we take a random sub-sample of $X_1,\ldots,X_M$ of length $m.$ So, we get sub-samples $X'^{(j)}_1,\ldots,X'^{(j)}_m,$ $j=1,\ldots,s.$
Now, we calculate the estimate of $D(X_i,\mathbb X)$ as
\begin{equation}
\label{eq:D1_approx}
D_{simpl}(X_i,\mathcal X)=\widehat{D}_{simpl}(X_i,\mathbb X):=\frac{1}{s}\sum\limits_{j=1}^{s}1(\exists L,U \in conv( X'^{(j)}_1,\ldots, X'^{(j)}_m):  \  L\subseteq X_i \subseteq U).
\end{equation}
Since $conv( X'^{(j)}_1,\ldots, X'^{(j)}_m)$ is infinite, in practice we cannot verify the above condition for all $U,L \in conv( X'^{(j)}_1,\ldots, X'^{(j)}_m).$ So, we choose $N \in \mathbb N$ and we can verify the condition only for $L,U \in \{ p_1X'^{(j)}_1+\ldots+p_mX'^{(j)}_m: p_l=\frac{n_l}{N}, n_l \in \mathbb N, \sum\limits_{l=1}^{m}n_l=N\}.$

\subsection{Comparison of depths}
\label{sec:comp}
The choice of the depth depends on the nature of the problem we are applying it to. 
Therefore here we explore the sensitivity of each proposed depth function to various features of sets such as the shape of the boundary and topological features (holes, connected components) when detecting the outliers within the sample.

For that purpose, we simulate the sample of 100 discs with the random radii from the uniform distribution on $[2,4]$  and add to the sample the sets from the first column of Table \ref{tab:Depth_comparison}: an ellipse with minor semi-axes of length 3.8 and mayor semi-axes equal 2.2, a square with sides length equal to 5, an annulus with the larger circle of radius 3 and the smaller one of radius 0.8, a disc of radius 3 with randomly excluded pixels, a union of a disc with a radius 2.5 and two small discs of radius 0.5, a union of a disc of radius 3  
 and a small disjointed disc with a radius of 0.3 and the union of two discs of with radii 2.8 and 1.5. All the sets were represented as matrices $100\times 100$ of zeros and ones. We calculate depths of the each of the added sets within the sample and observe which ones are recognised as the outliers, i.e. the empirical depth of the set within the sample is less than 0.05. The results are summarised in the remaining columns of the Table \ref{tab:Depth_comparison}.
 
When detecting the outliers using $D_{sign}$ we used function \verb|shape.fd.outliers| form package \verb|ddalpha| in R where second-order outliers are obtained as functions having second-order integrated depth smaller than 0.05 and differ from the sample in terms of growth. Third-order outliers are the functions different in convexity and concavity proprieties and are calculated as the outliers with respect to the third-order integrated depth. For more details the reader is referred to \cite{gijbels:2017}.
Band depth was approximated using $(\ref{eq:D_band_est})$ and setting $n=3.$ Depth $D_{simpl}$ estimated using (\ref{eq:D1_approx}) with $s=100,$ $m=3$ and $N=5.$

 We observe that the depth based on the signed distance function can  detect the outliers that are based on the differences in the shape of the boundary, 
 while the band depth and simplicial depth are better at detecting differences in the topology (holes). The other depths failed to recognise features of interest, so we do not consider them later in the paper.
 
 \begin{table}[H]
    \centering
    \begin{tabular}{|c||c|c|c|c|c|c|c|}
    \hline 
      & $D_{inf}$ & $D_{sign}$ & $D_{band}$ & $D_{simpl}$ &  $D_{ex}$ & $D_{hauss}$ & $D_{leb}$\\ \hline
         \includegraphics[scale=0.3]{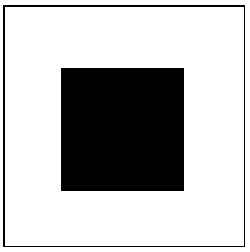}& - & + (second order) & - & - & - & - & -\\ \hline \hline
         \includegraphics[scale=0.3]{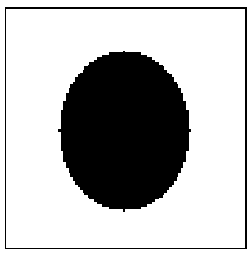} & - & + (second order) & -&- & - & - & - \\ \hline
         \includegraphics[scale=0.3]{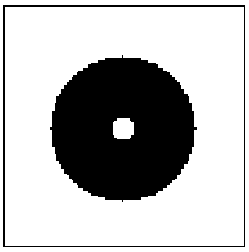}  & -& - &+ &+ &- &- &- \\ \hline
         \includegraphics[scale=0.3]{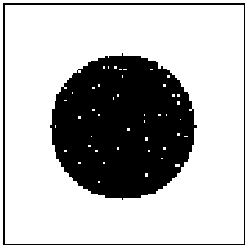} & - &+ (third order)  & + &+ &- &- & -\\ \hline
         \includegraphics[scale=0.3]{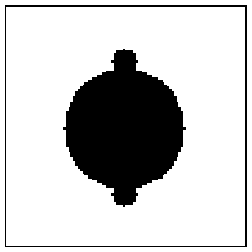} & -&  + (second order)& - & - &- &- & -\\ \hline
         \includegraphics[scale=0.3]{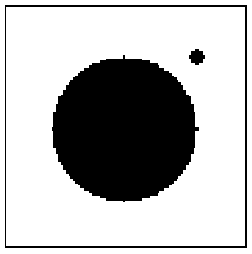} & -  & - & + & + & + & + & -\\ \hline
        \includegraphics[scale=0.45]{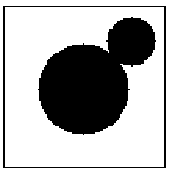} & -& - & + & + &  + & - & -\\ \hline
    \end{tabular}
    \caption{We simulated a sample of 100 discs of random radius and added to it the sets from the first column. The "+" means that the depth recognised the set as an outlier in the sample.}
    \label{tab:Depth_comparison}
\end{table}
\section{Application on comparing two samples of  random sets}
\subsection{DD plot}
\label{subsec:ddplot}
The notion of DD-plot was introduced in \cite{ddplot1}
for the purpose of comparing multivariate distributions of two samples using depth measures.

Let $\mathcal{X}=\left\{X_{1}, \ldots, X_{n}\right\}$ and $\mathcal{Y}=\left\{Y_{1}, \ldots, Y_{m}\right\}$ be the two samples of random sets we want to compare. The \textbf{DD-plot} of the combined sample is set of ordered pairs $DD(\mathcal{X}, \mathcal{Y})=\left\{\left(D(F,\mathcal X), D(F,\mathcal{ Y})\right), F \in \mathcal X \cup \mathcal Y \right\},$ where $D$ is an arbitrary depth function. So, DD-plot consists of $n+m$ ordered pairs of numbers between 0 and 1.

Since depths try to characterise the distributions of samples it is expected that DD-plots between equally distributed samples must be similar to the identity. Therefore, the scatter plot generated by the DD-plot should concentrate towards the $(0,0)-(1,1)$ line. 

The  DD-plot of samples with different distributions is dispersed and sometimes irregular. The nature of its irregularity can suggest different departures from equality in distribution (see \cite{ddplot1}).

To see how different parameters affect the shape of the DD-plot we use a random particle model generated in two steps. 

\begin{figure}[H]
    \centering
    \includegraphics{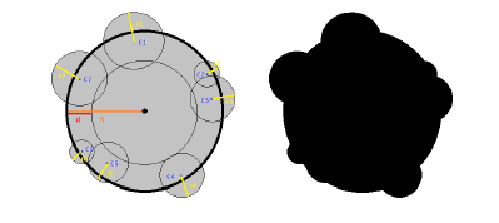}
   \label{fig:particle_model}
\end{figure}

\begin{figure}
    \centering
        \includegraphics{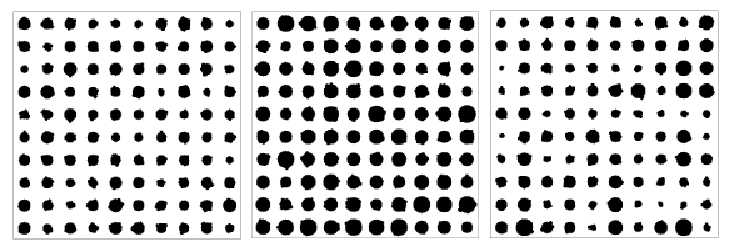}    
        \caption{Samples of size 100 generated from the random particle model with $(\mu_R,\sigma_R,d,\lambda,\mu_r,\sigma_r)=(3,0.5,9,1.5,5,3)$ (left) $(\mu_R,\sigma_R,d,\lambda,\mu_r,\sigma_r)=(\mathbf{12},1.5,5,3,3,0.5)$ (middle), $(\mu_R,\sigma_R,d,\lambda,\mu_r,\sigma_r)=(9,\mathbf{2.5},5,3,3,0.5)$ (right).}
    \label{fig:particle_samples}
\end{figure}

\begin{figure}[H]
    \centering
          \includegraphics[scale=0.8]{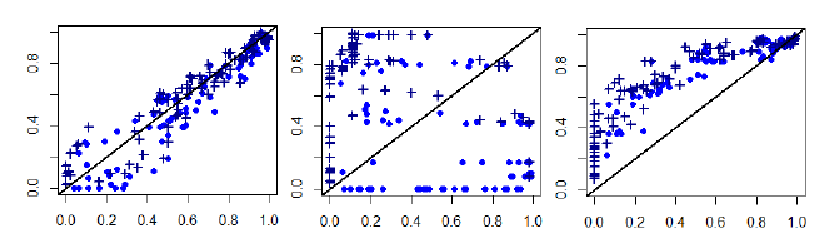}

        \caption{DD-plots obtained when comparing two samples from Model 1 (left), sample from Model 1 vs sample from Model 2 (middle), sample from Model 1 vs sample from Model 3 (right).}
    \label{fig:particle_ddplot}
\end{figure}

In the first step, the ball of a random radius $R$ is generated,  \break $R=\min\{ 0, N(\mu_R,\sigma_{R})\}$ where $N(\mu, \sigma)$ stands for a normal random variable with mean $\mu$ and standard deviation $\sigma.$ In the second step, for given $d \in \mathbb R, \ d>0$ and $N \sim P(\lambda)$ where $P(\lambda)$ is Poisson distribution with parameter $\lambda,$ $N$ points are independently and randomly chosen from the annulus (ring) within the ball generated in the first step with edges consisting of smaller ball with radii $R-d$ and larger ball with radii $R.$ These $N$ points are centres of balls with radii independently chosen from $\min\{0, N(\mu_r,\sigma_r)\}.$ The random particle is obtained as the union of the first ball with radius $R$ and the $N$ balls obtained in the second step. The graphical representation of the obtained particle model is given in Figure \ref{fig:particle_model}. 
We simulate 6 samples of 100 realisations from the above-mentioned model, 4 samples from Model 1  where $(\mu_R,\sigma_R,d,\lambda,\mu_r,\sigma_r)=(3,0.5,9,1.5,5,3),$ 1 sample Model 2 where $(\mu_R,\sigma_R,d,\lambda,\mu_r,\sigma_r)=(\mathbf{12},1.5,5,3,3,0.5)$ and  1 sample from Model 3 where $(\mu_R,\sigma_R,d,\lambda,\mu_r,\sigma_r)=(9,\mathbf{2.5},5,3,3,0.5)$   (see Figure \ref{fig:particle_samples}). Note that Model 1 and Model 2 differ in the location parameter of the radius of the first ball while Model 1 and Model 3 differ in the scale parameter of the radius of the first ball.
DD-plots based on the samples when comparing Model 1 vs Model 1, Model 1 vs Model 2 and Model 1 vs Model 3 are obtained using $D_{band}$ from (\ref{eq:D_band_est}) where $n=3$ and are visualised in Figure \ref{fig:particle_ddplot}.
We can observe that the DD-plot of two samples from the same model concentrates around the diagonal, while in the case of the difference in the centre is dispersed from the diagonal and in the case of the difference in the scale parameter of the radius of the "main" ball it is concentrated around a curve connecting (0,0) and (1,1).

\subsection{Testing equality in distribution of two samples of random  sets}

Authors of \cite{ddplottest} propose some statistics that can capture how concentrated the DD-plot is towards the $(0,0)-(1,1)$ diagonal. 

If the same process generated the two samples then the DD-plot should concentrate towards the $(0,0)-(1,1)$ diagonal line. For testing equality in the distribution of the two samples they propose to model the depths of two samples in the following way

\begin{equation}
\label{eq:reg}
D(F,\mathcal{Y})=\beta_{0}+\beta_{1} D(F,\mathcal X)+u_{F} \quad F \in \mathcal X \cup \mathcal Y.
\end{equation}
with $u_{F}$ being the  error. Under the null hypothesis $\beta_{0}=0$ and $\beta_{1}=1.$

The bootstrap-t procedure is used to check if $\beta_{0}=0$ and $\beta_{1}=1$ from which two $p$-values are obtained: $p_0$ and $p_1,$ respectively. 
For chosen confidence level $\alpha$, using the Holm-Bonferroni method, the $p$-values are sorted in ascending order, getting $p_{(1)}, p_{(2)}$. Finally, the null hypothesis is rejected if $p_{(1)}<\alpha/2$ or $p_{(2)}<\alpha.$ The adjusted $p$-value of the test is $p=\min \left\{2 p_{(1)}, p_{(2)}\right\}$.

 However, this test has low power in our case and fails to recognise some differences between the distributions of the samples. To describe where the above procedures are not satisfactory and the way to propose a new testing procedure, we introduce the following example.  
 
\begin{figure}[H]
     \centering
     \includegraphics{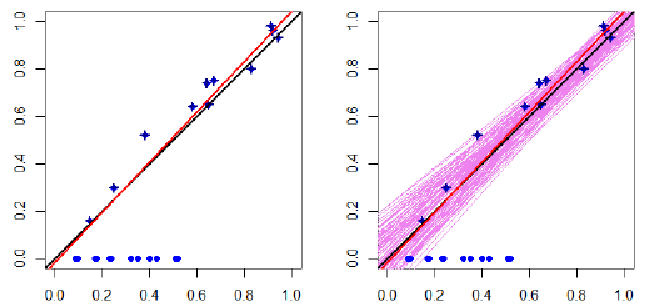}
     
     \caption{DD-plot obtained from the samples of size 100 of random sets: first one are random discs sample,  second sample is a mixture sample of random discs and random annulus. The dark blue crosses correspond to the points in DD-plot obtained from the depths of sets in the first sample, while the blue dots correspond to the pairs of depths of sets from the second sample. The red line represents the fitted regression line. The violet lines on the right plot are the regression lines obtained from the bootstrap procedure.}
     \label{Fig:counerex_ddplot}
 \end{figure}

 We generate two samples of size 100 of random sets. The first one consists of random balls with a radius uniformly distributed over interval $[8,10]$ and the second sample is a mixture sample, with probability 0.8 the realisation is a random ball with a radius uniformly distributed over interval $[8,10]$ and with probability $0.2$ the realisation is random annulus with smaller radius uniformly distributed over $[2,4]$ and larger radius uniformly distributed over interval $[8,10].$ The DD-plot obtained using $D_{band}$ is visualised in Figure \ref{Fig:counerex_ddplot}. The points in DD-plot originating from the first sample are denoted with dark blue crosses and the points belonging to the second sample are visualised as blue dots. From the DD-plot it is clear that the part of the samples have the same distributions since they fit the diagonal well. There are also a few blue dots (belonging to the rings in the second sample) having the first coordinate, i.e. the depth in the first sample, equal to 0 indicating that they are outliers with respect to the distribution of the first sample.
The above test gives $p$-values $p_0=0.65, \ p_1=0.38$,  although the samples significantly differ.
So, we propose to use a permutation version of the global envelope test (\cite{envelope}) where the test function is the difference between depths. Note that the difference between depth is proportional to the signed distance from the corresponding point in a DD-plot to a (0,0)-(1,1) line.
In more detail, our test function is 
\[
T_1(k)=D(F_k,\mathcal X)-D(F_k,\mathcal Y), \ k=1,\ldots,N+M,
\]
where $F_k$ stands for the $k-$th set in the joined sample $X_1,\ldots, X_N,Y_1,\ldots,Y_M.$

For chosen number of permutations $S,$ we permute joined sample $X_1,\ldots,X_{N},\break Y_1,\ldots, Y_M,$ and obtain $\mathcal X^{*}_i$ and $\mathcal Y^*_i,$ first sample consisting of first $N$ sets in the permuted sample and the second one consisting of the last $M$ sets in the permuted sample. We calculate $T_i(k)=D(F'_k,\mathcal X^*_i)-D(F'_k,\mathcal Y^*_i),$ $i=2,\ldots,S+1,$ where $F'_k$ is the $k-$th set in the permuted sample.
Under the assumption of the exchangeability of the joined sample, the distribution of  $T_i(k), i=2,\ldots,S+1$ should remain the same as the distribution of $T_1(k)$. 

In this way we obtain $S+1$ objects $T_1(k),T_2(k),\ldots, T_{S+1}(k),$ $k=1,\ldots, N+M.$

For each $k=1,\ldots, N+M,$ let $R_i^{\uparrow}(k)$ and $R_i^{\downarrow}(k)$ denote the ranks of the values $T^{(i)}(k)$ from the smallest value with rank $1$ to the largest one with rank $S+1$ and from the largest value with rank 1 to the smallest one with rank $S+1,$ respectively.
For each $k=1,\ldots, N+M,$ we define $k$-wise ranks of $T_i(k)$ as
$
R_i(k)=\min\left(R_i^{\uparrow}(k),R_i^{\downarrow}(k) \right).$
The extreme ranks $R_i$ are obtained by $R_i=\min_{k}R_i(k).$

In order to avoid the possibility of the ties, we use the area measure refinement of the extreme rank $R_i.$ It is constructed in the following way.

Let $T_{[1]}(k) \leq T_{[2]}(r) \leq \cdots \leq T_{[S+1]}(k)$ denote the ordered set of values $T_i(k), i=1, \ldots, S+1$. The continuous rank of $T_{[i]}(k)$ is
$$
c_{[i]}(k)=j+\frac{T_{[i]}(k)-T_{[i-1]}(k)}{T_{[i+1]}(k)-T_{[i-1]}(k)}, \quad \text { for } i=2, \ldots, S,
$$
and
$$
c_{[1]}(k)=\exp \left(-\frac{T_{[2]}(k)-T_{[1]}(k)}{T_{[S+1]}(k)-T_{[1]}(k)}\right), c_{[S+1]}(k)=S+1-\exp \left(-\frac{T_{[S+1]}(k)-T_{[S]}(k)}{T_{[S]}(k)-T_{[1]}(k)}\right) .
$$
Let 
$$
C_i(r)=S+1-c_i(k).
$$
Then the area rank measure $a_i$ defined as
$$
a_i=\frac{1}{S+1}\left(R_i-\frac{1}{N+M} \sum_k\left(R_i-C_i(k)\right) \mathbf{1}\left(C_i(k)<R_i\right)\right).
$$

Then the $p$-value of the test is the percentage of the curves having more extreme ranks than the observed one: 
\begin{equation*}
    p = \frac{1}{S+1}\left(\sum_{i=1}^{S+1} \mathbf{1}(a_i<a_1)\right).
\end{equation*}

If we want to get a graphical representation of the result, we proceed in the following way.  For a given significance level $\alpha$, we determine an appropriate  $a_{\alpha}$ 
which is the smallest rank in $\left\{  a_1,\ldots, a_{S+1}\right\}$
for which 
\begin{equation}
 \label{eq:pval}   
    \sum\limits_{j=1}^{S+1}1( a_j<a_{(\alpha)})\geq \alpha(S+1).
\end{equation}

Secondly, we construct the envelope of all the curves which have ranks smaller than $ a_{(\alpha)}.$
For
\[
    I_{\alpha}=\left\{
        j \in\left\{1,\ldots,s+1
        \right\}:
    a_j < a_{\alpha} \right\}
\]
we define values
$$
	T_{low}^{(\alpha)}(k) =
    \min_{j \in I_{\alpha}} T_j(k), 
    \quad T_{upp}^{(\alpha)}(k) = \max_{j \in I_{\alpha}}T_j(k).
$$ 

It holds
\[
    P\left(
        T_1(k) \notin
        \left[
            T_{low}^{\alpha}(k),T_{upp}^{(\alpha)}(k)
        \right]
        \text{ for any $k$} \left| H_0\right.
    \right)= \alpha.
\]

where $H_0$ is a simple null hypothesis.

Therefore, if 
the observed vector $T_1$ leaves the envelope at some point,  the null hypothesis is rejected at the significance level
$\alpha$.
If the observed vector lies completely inside this envelope, the null hypothesis is not rejected
at significance level $\alpha$.

\begin{figure}[H]
    \centering
     \centering
    \includegraphics{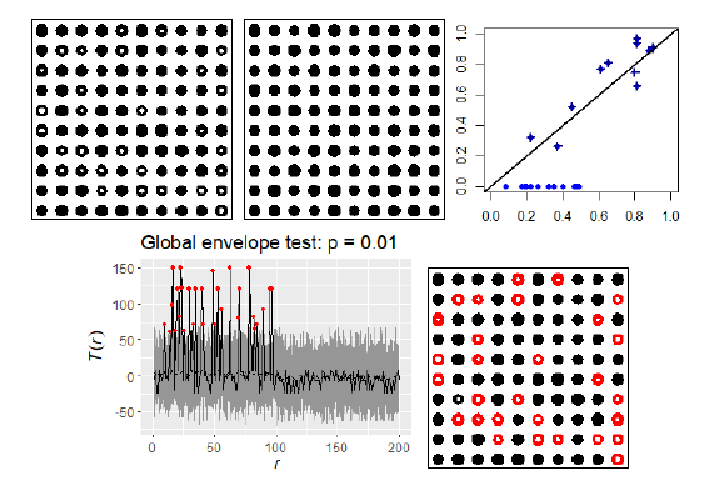}
    \caption{Second-row right plot: Results of the global envelope test when comparing the mixed sample with a sample of random balls. The result is obtained using package GET in R \cite{get}. Second row left plot: Mixed sample, the sets coloured in red are the ones for which the values of the difference between depths fall outside the 95\% envelope.}
    \label{fig:envelope}
\end{figure}

We create a $95\%$ envelope and in this way, if the null hypothesis is not accepted, we can see which sets (the arguments $k$ such that test function $T(k)$ falls outside the envelope) are responsible for the rejection (see left plot of Figure \ref{fig:envelope}). The sets responsible for the rejection belong to the first sample and were coloured red in the right plot of Figure \ref{fig:envelope}.

\section{Simulation study}
The first aim of the simulation study is to compare the test based on DD-plots introduced in \cite{ddplottest} with the test proposed in this paper.
For that purpose, we use a random particle model explained in Section \ref{subsec:ddplot}.

\begin{figure}[H]
    \centering
    \includegraphics[height=4cm,width=4cm]{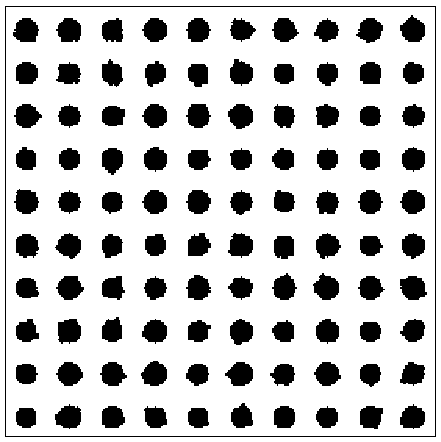}
    \includegraphics[height=4cm,width=4cm]{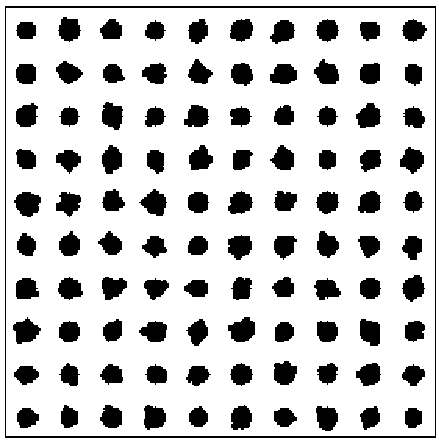}

    \caption{Sample from Model 4 (left) and sample from Model 5 (right)}
    \label{fig:particles}
    \end{figure}

Two samples of random particles generated from this model using different parameters are shown in Figure \ref{fig:particles}. For the left figure, the parameters were
$(\mu_R,\sigma_R, \lambda, d, \mu_r,\sigma_r)=(9,0.2,5,3,4,0.5)$ and will be further referred as Model 4. The right figure represents the sample from the random particle model referred to as Model 5 with the parameters $(\mu_R,\sigma_R, \lambda, d, \mu_r,\sigma_r)=(10,1,4,2,3,0.2).$

\begin{figure}[H]
    \centering
     \includegraphics[scale=0.6]{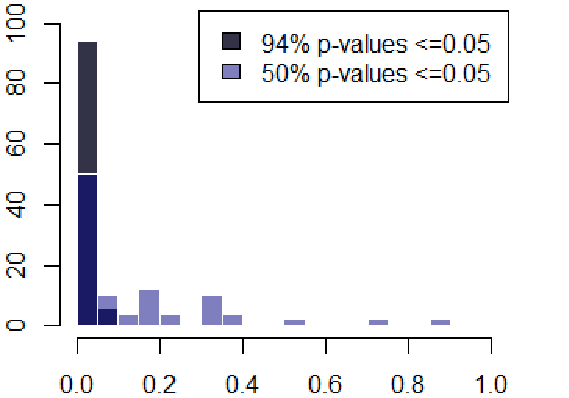} 
    \caption{Histogram of the $p$-values of test when comparing Model 4 and Model 5  when depths are evaluated using (\ref{eq:D_band_est}) with $n=5,$ $s=100$ and $S=100$ when using test from \cite{ddplottest} (light blue) and newly introduced test based on the global envelope test (dark blue). }
    \label{fig:hist_Uz1_Uz2}
    \end{figure}

The histograms of $p$-values obtained when comparing Model 4 with Model 5 are presented in Figure \ref{fig:hist_Uz1_Uz2}. It is clear that the newly proposed testing procedure has a greater power compared to a testing procedure proposed in \cite{ddplottest}.
The histograms of $p$-values when comparing the same models are approximately uniformly distributed in both cases.

To see how the DD-plots can depict the difference between distributions of random sets and to compare the power of the test introduced in the previous section with the permutation test based on $\mathfrak N$-distances introduced in \cite{gotovac:2019} we first use realisations of 4 simulated processes as in \cite{gotovac:2019}. The idea is to test whether the connected components of the realisations are equally distributed and if not find possible clues what are the differences in their distributions. 

\begin{figure}[H]
 \centering
 \includegraphics[]{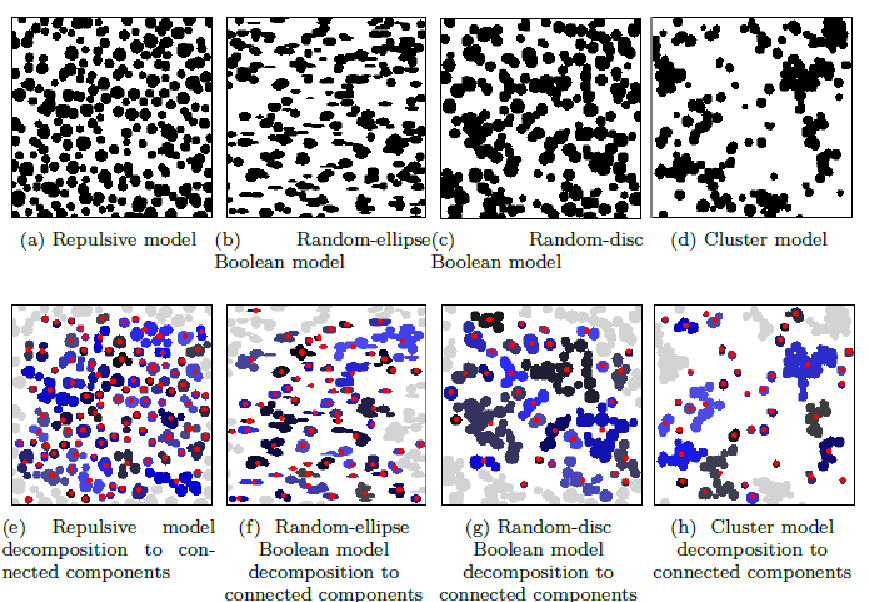}
\caption{Examples of realisations of the simulated repulsive model  (a) and its decomposition to connected components  (e), random-ellipse Boolean model (b) and its decomposition to connected components (f), random-disc Boolean model (c) and its decomposition to connected components  (g) and cluster model (d) and its decomposition to connected components  (h). Figures are taken from \cite{gotovac:2019}.}
\label{fig:models}
\end{figure}

 The first model is a random-disc Boolean model with centres of discs in the window $25 \times 25$, the intensity of the disc centres equal to $0.4$ and the uniform distribution of radii on the interval $(0.5, 1)$ (see Figure \ref{fig:models} (c)).  The second one is the random-ellipse Boolean model with centres of ellipses in the window $25 \times 25$, the intensity of the ellipse centres equal to $0.4$ and uniform distribution of semi-major axes on the interval $(0.5, 1)$ and semi-minor axes on interval $(0.2,0.7)$ (see Figure \ref{fig:models} (b)).
  The third one, referred to as a cluster process, is the Quermass-interaction process  (see) with the parameters  $\theta_1$ = 0.62, $\theta_2$ = -0.86 and $\theta_3$ = 0.7 with respect to the random-disc  Boolean model mentioned above. Its realisations have a larger area and smaller perimeter compared to the reference process, it tends to create clusters (see Figure \ref{fig:models} (d)). The fourth model referred to as a repulsive model, is simulated as a Quermass-interaction process with parameters $\theta_1$ = -1, $\theta_2$ = 1 and $\theta_3$ = 0 with respect to the same random-disc   Boolean model. The realisations have a smaller area and larger perimeter than the reference random-disc   Boolean modes, so they consist of small non-overlapping components (see Figure \ref{fig:models} (a)).

  We have simulated $200$ realisations for each of the mentioned processes, all realisations are transformed to matrices of 400 $\times$ 400 black and white pixels.
  For each realisation, the connected components were isolated and the ones that were not completely visible within the observation window (i.e. the ones touching the boundary) were removed. The remaining connected components were centred around their centres of mass (see Figure \ref{fig:models} (e)-(h)) and formed samples on which we work.

  When comparing two processes, 100 different pairs consisting of one realisation from each of the compared processes were considered.

  We considered band depth, simplicial depth and depth based on signed distance function representation with the usage of the second-order extended integrated depth for functional data.
  The reason for choosing these depths is that in Section \ref{sec:comp} they have shown to recognise features we want to concentrate on such as the shape of the boundary of the components, their size and topology.
  The depth based on the expectation of random sets was not used since it obtains only values in the set $\{\frac{1}{m}: m=1,\ldots,n\}.$ It is possible for a set that is deep in both samples to have an empirical depth of 1 in one sample and an empirical depth of 0.5 in the second sample. Since in our testing procedure, the difference between two empirical depths plays the role of the test statistics, the testing procedure would recognise the difference between obtained empirical depths as significant, which is not desirable.

\begin{figure}[H]
    \centering
\includegraphics[scale=0.7]{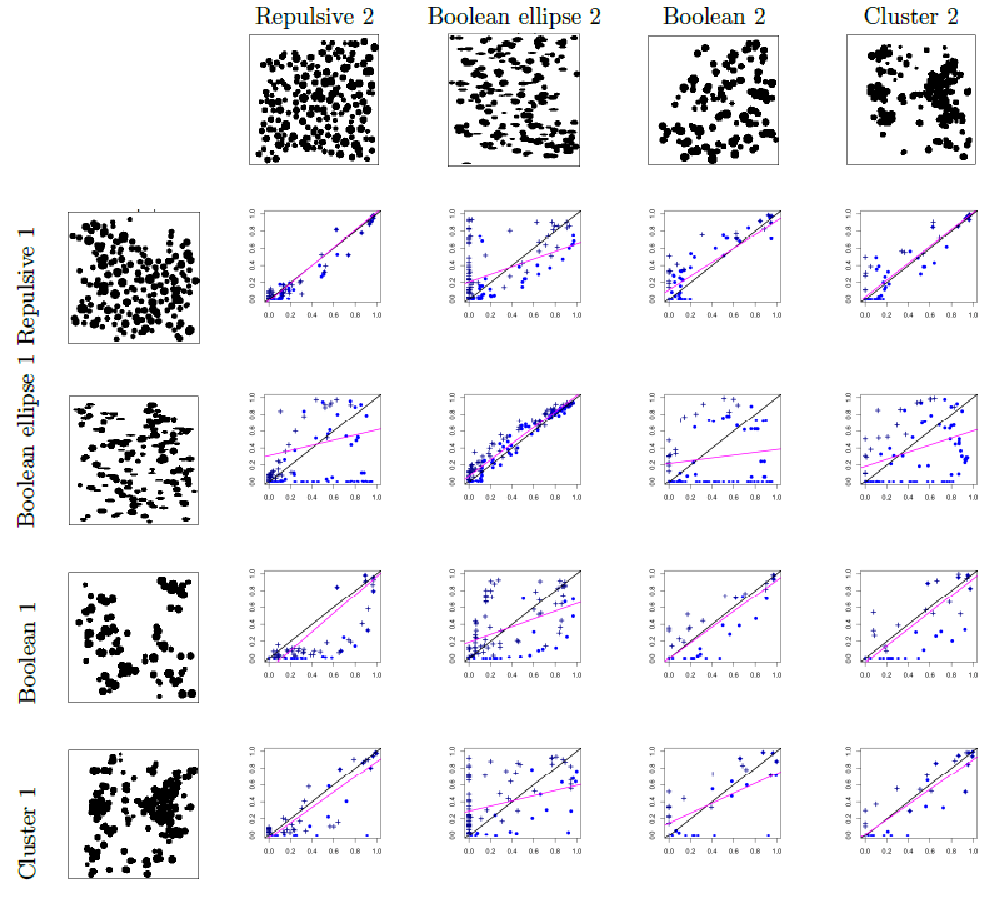}
    \caption{DD plots obtained when comparing different realisations. The first row and the first column represent realisations, while the remaining plots are DD-plots obtained when comparing the realisation at the beginning of the row with the realisation at the beginning of the column. The blue dots in the DD-plot originate from the first sample set's depths and the dark blue crosses correspond to the pairs of depth from the sets in the second sample. The violet line is the regression line for the points in the DD-plot.. The parameters used to approximate depth form equation (\ref{eq:D_band_est}) were $n=8$.}
    \label{fig:DDplots_BCR}
\end{figure}

\begin{figure}[H]
 \centering
\includegraphics[scale=0.7]{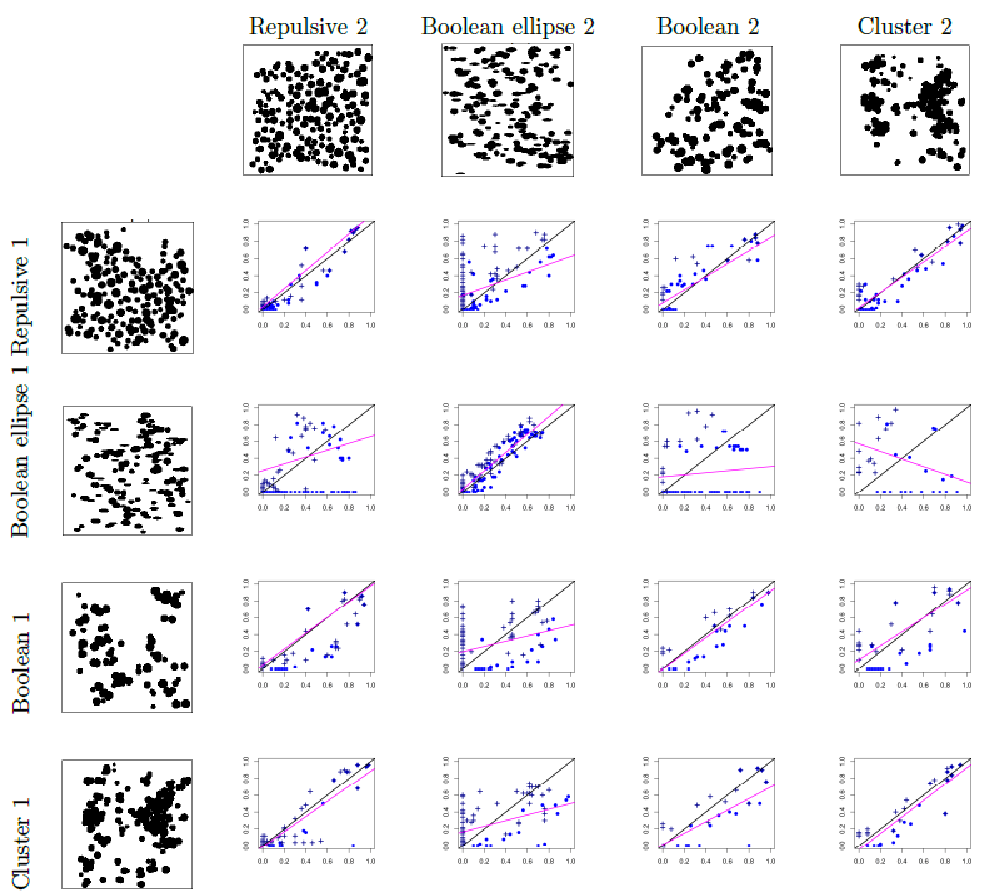}
    \caption{DD plots obtained when comparing different realisations. The first row and the first column represent realisations, while the remaining plots are DD-plots obtained when comparing the realisation at the beginning of the row with the realisation at the beginning of the column. The blue dots in the DD-plot originate from the first sample set's depths and the dark blue crosses correspond to the pairs of depth from the sets in the second sample. The violet line is the regression line for the points in the DD-plot.  The parameters used to approximate depth form equation (\ref{eq:D_simp}) were $m=5$ and $s=100$ and $N=5$. }
    \label{fig:DDplots_simp_s_BCR}
\end{figure}

\begin{figure}[H]
 \centering
\includegraphics[scale=0.7]{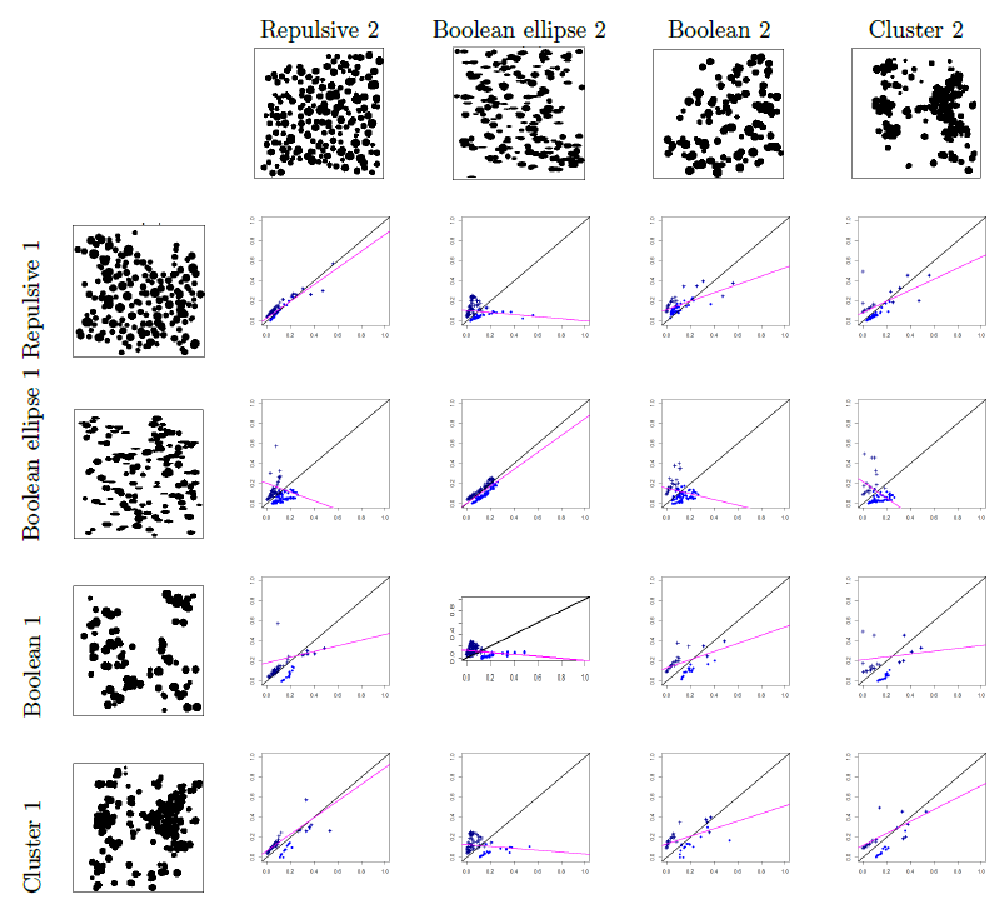}
    \caption{DD plots obtained when comparing different realisations when using $D_{sign}$. The first row and the first column represent realisations, while the remaining plots are DD-plots obtained when comparing the realisation at the beginning of the row with the realisation at the beginning of the column. The blue dots in the DD-plot originate from the first sample set's depths and the dark blue crosses correspond to the pairs of depth from the sets in the second sample. The violet line is the regression line for the points in the DD-plot.  
    }
    \label{fig:DDplots_BCR_D_sign}
\end{figure}

  The DD-plots obtained when comparing pairs of realisations of aforementioned processes are presented in Figures \ref{fig:DDplots_BCR}, \ref{fig:DDplots_simp_s_BCR} and \ref{fig:DDplots_BCR_D_sign}, using band depth, simpical depth and depth based on signed distance function representation, respectively.

  We estimated band depth using Equation (\ref{eq:D_band_est}).
  Parameter $n$ was chosen as the smallest one for which all values from 0 to 1 are observed in an empirical depth and $s=1000.$
  The simplical depth was estimated using $(\ref{eq:D1_approx})$ with parameters $s=1000,$ $m=5$ and $N=5.$
  
  Calculation of band depth and simplicial depth for $S=99$ permuted samples was executed programming language Julia \cite{julia} since it was much faster than in R. The value of $S$ was chosen to be the largest value such that the calculations could be executable in real-time. The envelope test was performed in R using function \verb|global_envelope_test| from package \verb|GET|.

  The depth based on the signed distance function was calculated in R using functions \verb|distmap| form package \verb|spatstat| for approximation of the signed distance function of and \verb|depthf.fd1| from package \verb|ddalpha| for calculating second-order extended integrated depth for both original and permuted samples.

We can see from Figure \ref{fig:DDplots_BCR} that the DD-plots obtained when comparing the realisations from the same model (the DD-plots on the diagonal) are concentrated towards the (0,0)-(1,1) line. The deviation from the (0,0)-(1,1) line is larger in the case of smaller samples. 

When comparing realisations from the repulsive and Boolean ellipse model, we can observe from the DD-plot that there are connected components in the Boolean ellipse model that have 0 depth in a sample of a repulsive model and the remaining components (more circular ones) resemble the repulsive model components. 
\begin{figure}[H]
\centering
\includegraphics[scale=0.8]{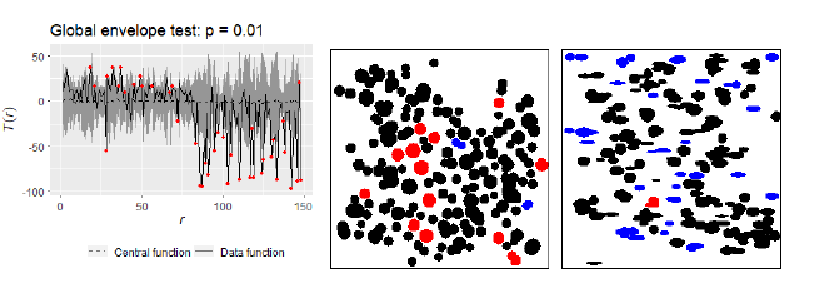}
\caption{First row: Results of the global envelope test when comparing Repulsive 1 with Boolean ellipse 2 (S=100). Second row: The components coloured in red are the ones that have fallen out of the 95\% envelope and have higher depth in the sample Repulsive 1 than in sample Boolean ellipse 2, while the components coloured in blue have fallen out of the 95\% envelope  but have significantly higher depth in sample Boolean ellipse 2 than in sample Repulsive 1.}
\end{figure}
When comparing the Repulsive and Boolean models using band depth and simplicial depth we observe from the DD-plots that the distribution of components has the same location parameter value, however, the scale of the Boolean component's distribution is larger.

Further on, we used the envelope test to test equality in the distribution of the components. The $p$-values obtained by $(\ref{eq:pval})$ when the same processes were compared were uniformly distributed, while when comparing different processes the $p$-values were smaller and closer to zero. The percentage of the $p$-values smaller than 0.05 when comparing different processes using different depths and using the test from \cite{gotovac:2019} were summarised in Table \ref{tab:mp-val_comparison}. From it, we see, that the test based on the band depth has the greatest power among the tests using depths. Test from \cite{gotovac:2019} has shown greater power when comparing all pairs of processes except the comparison of random-disc Boolean with cluster process. In this case, the test based on the depths performed better.

  \begin{table}[H]
      \centering
      \begin{tabular}{|c|c|c|c|c|c|c|}
      \hline
           &  R vs B & R vs Be & R vs C & B vs Be & B vs C & Be vs C\\
           \hline
           $D_{fun}$ & 6 & 100 & 12 & 54 & 20 & 54\\
           \hline
           $D_{band}$ & 42 & 100 & 44 & 98 &  54 & 94\\
           \hline
           $D_{sim}$ & 26 & 94 & 22 & 82 & 34 & 94\\
           \hline
           N-test & 95 & 100 & 49 &100 & 26 & 100\\
           \hline
      \end{tabular}
      \caption{The percentages of the $p$-values smaller than 0.05 when comparing repulsive with random-disc Boolean model (R vs B), repulsive with random-ellipse Boolean model (R vs Be), repulsive with cluster (R vs C), random-disc Boolean with random-ellipse Boolean (B vs Be), random disc Boolean with cluster (Be vs C) and random-ellipse Boolean with cluster model. The first row corresponds to the percentages of $p$-values smaller than 0.05 obtained when using depth based on the signed distance function, the second row when using band depth, the third when using simplicial depth and the fourth when using permutation test based on $\mathfrak N$-distances from \cite{gotovac:2019}. }
      \label{tab:mp-val_comparison}
  \end{table}
  \section{Application to real data}
\begin{figure}[H]
\centering
\includegraphics[scale=0.8]{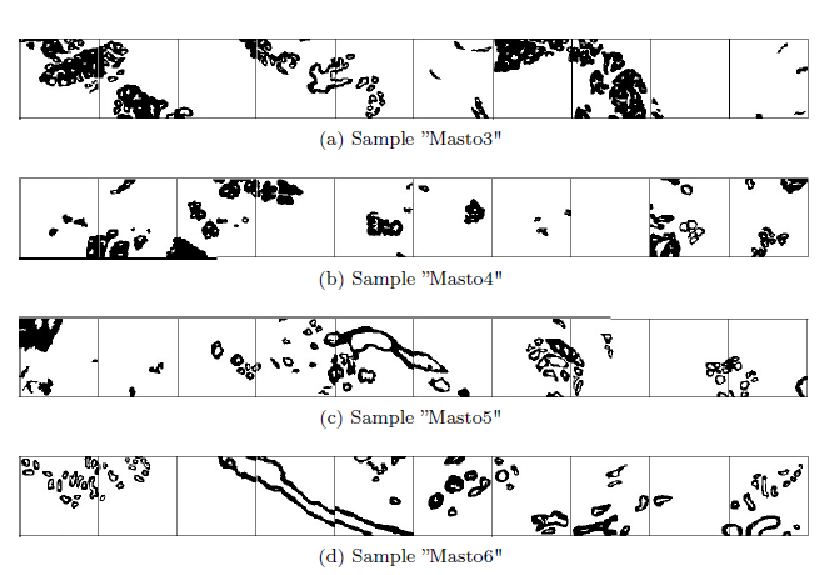}
\caption{Samples of masthopatic breast tissue kindly provided by (Mrkvi\v{c}ka T. and Mattfeldt T.(2011))}
\label{fig:mas}
\end{figure}

\begin{figure}[H]
 \centering
    \includegraphics[scale=0.8]{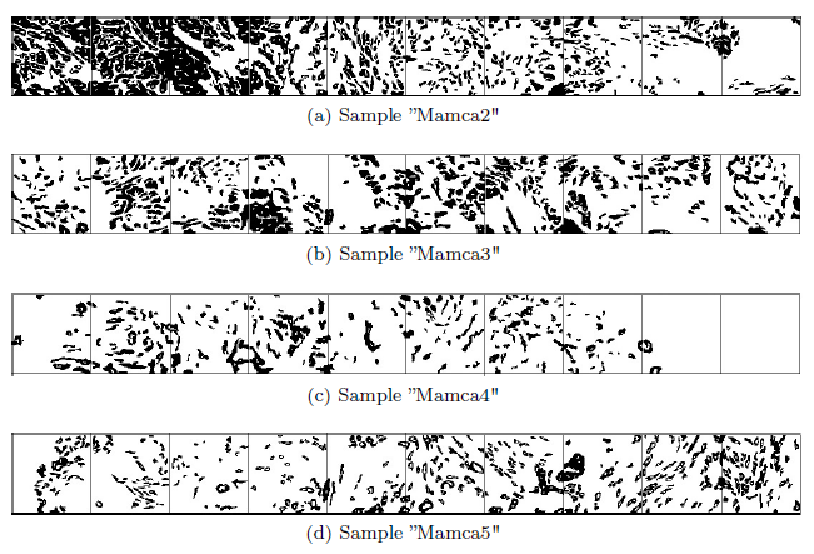}
\caption{Samples of mammary cancer kindly provided by (Mrkvi\v{c}ka T. and Mattfeldt T.(2011))}
\label{fig:mam}
\end{figure}

  In this section, we briefly present the results of comparing the 4 histological images of  mastopatic breast tissue (see Figure \ref{fig:mas}) with 4 histological images of mammary cancer tissue (see Figure \ref{fig:mam}) kindly provided by the authors of \cite{mrkvicka:2011}. The same images were considered in \cite{gotovac:2019}, here we use a subset of images for the reason of the better visualisation of the results in terms of tables with DD-plots. As in \cite{gotovac:2019}, each image was decomposed in a way that black pixels were first partitioned in the connected components and then each component was decomposed in a way that pixels were assigned to the closest hole which usually represents the ducts in the breasts. The idea was to capture the shape of the tissue surrounding the ducts. In that way, we obtained 8 samples of sets from each of the realisations. The samples were compared using a test based on band depth described in this paper since band depth has shown the best power in the simulation study. The depth was estimated using (\ref{eq:D_band_est}) with $n=10$ and $s=100.$ The envelope test was performed with $S=49$ number of permutations of the joined sample.
  
  The results when comparing mastopatic tissue samples are depicted in Table $\ref{tab:mas_mas}$. A comparison of mammary cancer tissue samples can be found in Table \ref{tab:mam:mam}. We can see that DD-plots when comparing samples from the same group are concentrated around the (0,0)-(1,1) line and the $p$-values of the test are in most cases significantly larger than 0.05. 
  
  Table \ref{tab:mam_mas} summarises the comparison between different processes, i.e. when comparing mastopatic tissue samples with mammary cancer tissue samples. The DD-plots are irregular but seem to have the same pattern of iregularity. Also, the $p$-values of the test are smaller than 0.05. Therefore, the test recognises the difference between the groups.

  \begin{table}[H]
  \centering
 \includegraphics[scale=0.8]{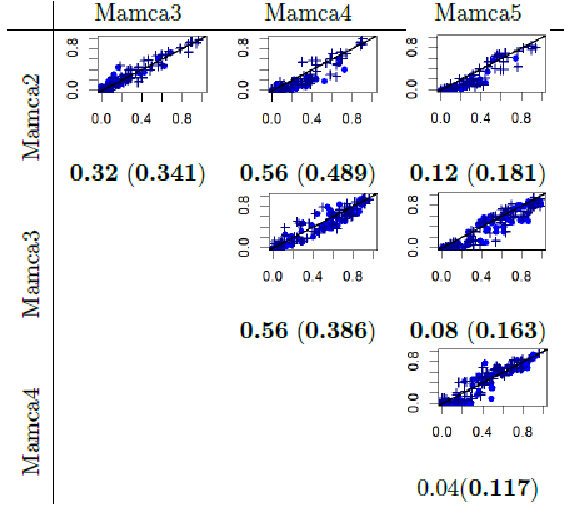}
\caption{DD-plots of samples of mastopathic tissue cancer when using band depth the $p$-values of the test based on depth when using envelope test and the results of the test based on $\mathfrak N$-distances form \cite{gotovac:2019} in the brackets.}
\label{tab:mas_mas}
\end{table}
\begin{table}
\centering
\includegraphics[scale=0.8]{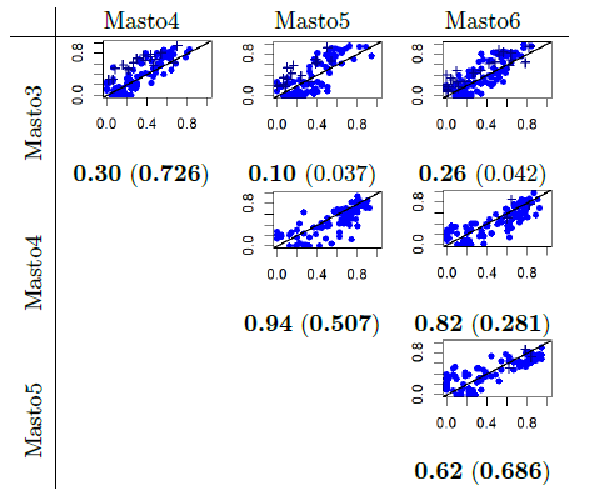}
\caption{DD-plots of samples of mammary cancer when using band depth, the $p$-values of the test based on depth when using envelope test and the results of the test based on $\mathfrak N$-distances form \cite{gotovac:2019} in the brackets. }
\label{tab:mam:mam}

\end{table}

\begin{table}[H]
\centering
\includegraphics[]{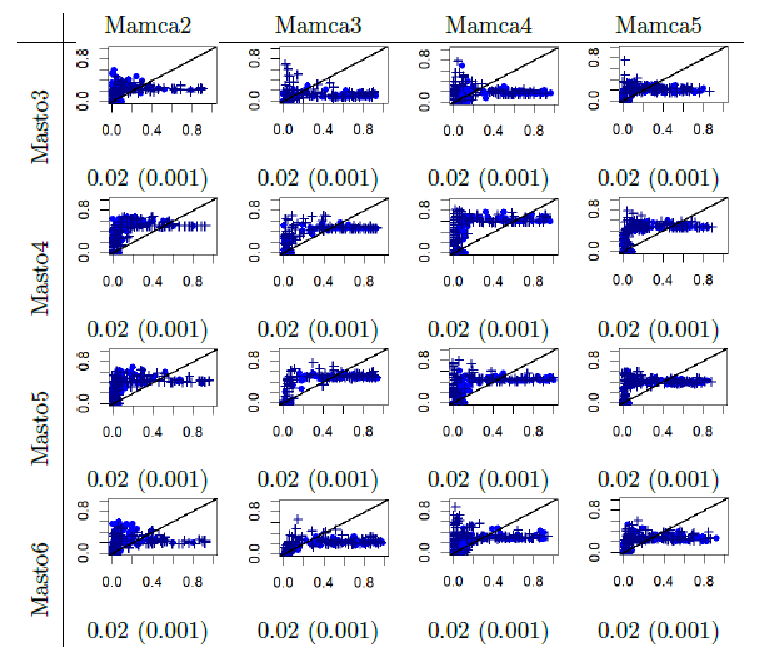}
\caption{DD-plots of samples of mammary cancer tissue and mastopatic breast tissue when using band depth, the $p$-values of the test based on depth when using envelope test and the results of the test based on $\mathfrak N$-distances form \cite{gotovac:2019} in the brackets.}
\label{tab:mam_mas}
\end{table}

\section{Discussion}
We have summarised various ways of defining the statistical depths for random non-convex sets. The estimation based on the samples was proposed. The properties of depths were discussed. Affine invariance guarantees us the same results in the case of the affine transformation of the data, while the upper semicontinuity assures small differences in the depth under small errors in the data.
The depths were compared to see the sensitivity to different shape features of the sets when detecting outliers. Band depth and simplicial depth have shown good results in detecting the outliers based on the size and the topology and the depth based on the signed distance function when using second-order integrated functional depth detected differences in the shape of the boundary.

The new test for testing equality in the distribution of random sets was proposed based on the DD-plots obtained from estimated depths. 

The simulation study showed that the proposed test based on DD-plot has lower power than the permutation test based on $\mathfrak N$-distances introduced in \cite{gotovac:2019} in all cases except when comparing random-disc Boolean with the cluster model and higher power than test proposed in \cite{ddplottest}. However, the newly introduced test brings valuable insights into the reasons for the rejection of the null hypothesis and together with the shape of the DD-plot can give possible clues on the differences between distributions of two samples.

Also, the proposed test using the band depth successfully classified the real data histological images of mammary cancer tissue and mastopatic breast tissue. Therefore, this method can be used for purposes of detecting the differences between distributions of the shapes of tissue.

\section*{ACKNOWLEDGEMENT}
\small
Supported in part by  the Croatian-Swiss Research Program of the Croatian Science Foundation and the Swiss National Science Foundation: project number IZHRZ0\_180549.

\end{document}